\def\tPm{\tilde{\bm P}_m}
\def\tPs{\tilde{\bm P}_s}

\def\Pm{{\bm P}_m}
\def\Ps{{\bm P}_s}
\def\Pns{{\bm P}_{\underline s}}
\def\eps{\varepsilon}
\def\half{\tfrac{1}{2}}
\def\trho{\tilde{\rho}}
\def\us{{\underline{s}}}

\newcommand{\vectornorm}[1]{\left|\left|#1\right|\right|}

\documentclass[twocolumn,osajnl2,showpacs,preprintnumbers,floats]{revtex4}
\usepackage {graphicx}
\usepackage {amsmath}
\usepackage {bm}
\usepackage {subfigure}

\def\@dotsep{4.5}
\makeatother

\begin{document}

\title{Phase retrieval and saddle-point optimization}

\begin{abstract}
Iterative algorithms with feedback are amongst the most powerful and
versatile optimization methods for phase retrieval. Among these, the
hybrid input-output algorithm has demonstrated practical solutions 
to giga-element nonlinear phase retrieval problems, escaping local minima
and producing images at resolutions beyond the capabilities of
lens-based optical methods. Here the input-output iteration is
improved by a lower dimensional subspace saddle-point optimization.
\end{abstract}

\author{Stefano Marchesini}
\altaffiliation[Current address: ]{Lawrence 
Berkeley National Laboratory, 1 Cyclotron Rd, Berkeley CA 94720, USA. 
e-mail: smarchesini@lbl.gov}
%\email[e-mail: ]{smarchesini@lbl.gov}
 \affiliation{Lawrence Livermore National Laboratory, 7000 East Ave.,
Livermore, CA 94550-9234, USA}
 \affiliation{
 Center for Biophotonics Science and Technology, University of California, Davis, 2700 Stockton Blvd., Ste 1400, Sacramento, CA 95817, USA}

%\pacs{61.10.Nz  42.30.Rx  42.30.Wb  68.37.Yz}

\preprint{UCRL-JRNL-226292}

%\date{\today}
\maketitle
%%%%%%%%%%%%%%%%%%%%%
\section{Introduction}
Phase retrieval is one of the toughest challenges in optimization,
requiring the solution of large-scale, nonlinear, non-convex and
non-smooth constrained problems.  Despite such challenge, efficient
algorithms are being used in astronomical imaging, electron
microscopy, lensless x-ray imaging (diffraction microscopy) and x-ray
crystallography, substituting lenses and other optical elements in the
image-forming process.

In diffraction microscopy, photons scattered from an 
object (diffraction pattern) are recombined 
solving the giant puzzle of placing millions of waves into a limited
area.  X-ray diffraction microscopy\cite{Miao:1999} has successfully 
been applied to image objects as complex as biological
cells \cite{shapiro:PNAS05}, quantum dots \cite{Miao:2006},
nanocrystals \cite{Pfeifer:2006} and nanoscale aerogel structures
\cite{barty}. Nanofabricated test objects were reconstructed
computationally in 3D with several millions 10 nm resolution elements 
\cite{chapman:JOSAA06}, other test patterns were captured in the 
fastest flash image ever recorded at suboptical resolution \cite{chapman:NP06}.

These experimental methods (see e.g. \cite{spence:book} for a review)
are being developed thanks to advances in optimization techniques,
primarily the introduction of a control-feedback method proposed by
Fienup (Hybrid Input Output-HIO \cite{Fienup:1978,Fienup:1982}).  The
important theoretical insight that these iterations may be viewed as
projections in Hilbert space \cite{stark:1984,stark:1987} has allowed
theoreticians to analyze and improve on the basic HIO algorithm
\cite{elser:2003, luke:1,luke:2,luke:3}.  
 More recently Elser et
al. \cite{elser:PNAS07,elser:PRE} connected the phase retrieval problem to other
forms of ``puzzles'', demonstrating performances of their Difference
Map algorithm\cite{elser:2003} (a generalization of HIO) often superior to dedicated
optimization pakages in problems as various as graph coloring, protein
folding, sudoku and spin glass states. 
Rather than performing a local
optimization of a constrained problem, the common theme of these
algorithms is that they seek a solution to a different type of fixed
point, the saddle-point of the difference between antagonistic error
metrics \cite{unified} with respect to the feasible and unfeasible
spaces defined by the constraints.

Here each input-output iteration is improved by a lower dimensional
subspace optimization of this saddle-point problem along the steepest
descent-ascent directions defined by the constraints.  This lower
dimensional optimization (performed here by Newton methods) is
analogous to one dimensional line searches of gradient based
methods, used to avoid overshooting and undershooting in new search
directions and providing faster and more reliable algorithms.

The first sections introduce the phase retrieval problem and the
saddle point optimization method, reformulating the HIO algorithm in
terms of gradients and constraints (see \cite{unified} for further
details).  Sections \ref{sec:2D} describe this lower dimensional
optimization. Benchmarks performed on
a simple simulated test pattern are described in the Sec. \ref{sec:performance}.

\section{Phase retrieval problem}
When we record the diffraction pattern intensity of light scattered by
an object, the phase information is lost.  Apart from normalization
factors, an object with density $\rho(\bm{r})$, $\bm{r}$ being the
coordinates in the \textit {object} (or \textit {real}) space,
generates a diffraction pattern intensity equal to the modulus square
of the Fourier Transform (FT) $\trho(\bm{k})$:
\begin{eqnarray}
I(\bm{k})&=&|\trho(\bm{k})|^2 %\\ %I(\bm{k})&=&\trho^\dagger (\bm{k}) \trho (\bm{k})\,,
\label{eq:phase}
\end{eqnarray}
where $\bm k$ represent the coordinate in the Fourier (or Reciprocal)
space. In absence of constraints, any phase $\varphi(\bm k)$ can be
applied to form our solution $\trho=\sqrt{I}e^{i \varphi}$. 

Phase retrieval consists in solving (Eq. (\ref{eq:phase})) from the measured
intensity values $I(\bm k)$ and some other prior knowledge
(constraints). Diffraction microscopy solves the phase problem using the
knowledge that the object being imaged is isolated, it is assumed to
be 0 outside a region called support $S$:
\begin{equation}
\rho(\bm{r})= 0  \text {,   if  $\bm{r}\notin S$}. 
\label{eq:support}
\end{equation}
In practical experiments, where the object interacts by absorbing as
well as refracting incident light, the problem is generalized 
to the most difficult case of objects with complex ``density''
or index of refraction.  In case of complex objects, this ``support''
region is the only constraint and needs to be well defined, tightly
wrapping the object. In many cases high contrast, sharp objects or illuminating beam
boundaries are sufficient to obtain such support of the object
ab-initio \cite{Marchesini:2003, chapman:JOSAA06}.

A projection onto this set ($\Ps$) involves setting to 0 the components
outside the support, while leaving the rest of the values unchanged:
\begin{equation}
\bm{P}_s \rho(\bm{r})=
\begin{cases}
\rho(\bm{r}) &\text {if  $\bm{r}\in S$}, \\
 \\
0 &\text{otherwise}.
\end{cases}
\end{equation}
Its complementary projector can be expressed as 
$\Pns={\bm I}-\Ps$.

The projection to the nearest solution of (Eq. (\ref{eq:phase}))
in reciprocal space is obtained by setting the modulus to the measured one  
$m(\bm k)=\sqrt{I(\bm k)}$, and leaving the phase unchanged:
\begin{eqnarray}
\label{eq:modulus}
\tPm \trho(\bm{k})&=&
\tPm |\tilde \rho(\bm{k})|e^{i\varphi(\bm{k})}=\sqrt I
(\bm{k})e^{i\varphi(\bm{k})}\,,
\end{eqnarray}
Singularities arise when $\trho$ is close to 0, and a small change in
its value will project on a distant point. Such projector is a 
``diagonal'' operator in Fourier space, acting 
element-by-element on each amplitude. When applied to 
real-space densities $\rho(\bm r)$, it becomes non-local, mixing
every element with a forward
${\cal F}$ and inverse ${\cal F}^{-1}$  Fourier transform:
\begin{eqnarray}
\Pm &=&{\cal F}^{-1} \tPm {\cal F}\,.
\end{eqnarray}
The Euclidean length $||\rho||$ of a vector $\rho$ is defined as:
\begin{equation}
||\rho||^2=\rho^{\dag} \cdot \rho=\sum_{\bm r} |\rho(r)|^2=
\sum_{\bm k} |\tilde \rho(k)|^2.
\end{equation}
If some noise $\sigma(k)$ is present, the sum should be weighted by
$w=\tfrac{1}{\sigma^2}$. 
The distance from the current point to the corresponding set 
$||\bm P \rho -\rho||$ is the basis for our error metrics: 
\begin{eqnarray}
\nonumber
\eps_s(\rho)&=&\vectornorm{\bm{P}_s \rho-\rho}, \\
\eps_m(\rho)&=&\vectornorm{\bm{P}_m \rho-\rho},
\end{eqnarray}
or their normalized version $\overline
\eps_{s,m}(\rho)=\tfrac{\eps_{s,m}(\rho)}{\vectornorm{\bm{P}_{s,m} \rho}}$.

 The gradients of the squared error metrics can be expressed in 
terms of projectors\cite{Fienup:1982,luke:siam}:
\begin{eqnarray}
\label{eq:gradient_m}
\nabla\eps^2_m(\rho)&=&-2[\bm{P}_m-\bm{I}]\rho\\
\nabla\eps_s^2(\rho)&=&-2[\Ps-\bm{I}]\rho\,,
\label{eq:gradient_s}
\end{eqnarray}
Steps of $-\half \nabla \eps^2_{s,m}$ bring the corresponding
error metrics to 0. The solution, hopefully unique, is obtained when both
error metrics are 0.

\section{Minimization in feasible space \label{sec:ER}}
One popular algorithm \cite{Gerchberg:1972,Fienup:1982} fits the data by
minimizing the error metric $\varepsilon_m(\rho)$:
\begin{eqnarray}
&&\min_{\rho} \varepsilon_m^2(\rho)\,,\\
\nonumber
&&\text{subject to $\Pns \rho=0$,}
\end{eqnarray}
by enforcing the constraint and moving only in the feasible
space $\rho_s=\Ps \rho$. The problem is transformed into an unconstrained
optimization with a reduced set of variables $\rho_s$:
\begin{eqnarray}
\min_{\rho_s} \varepsilon_m^2(\rho_s)\,.
\end{eqnarray}
The steepest descent direction is  projected in the feasible space:
\begin{eqnarray}
\label{eq:ER}
\nonumber
\rho^{(n+1)}&=&\rho^{(n)}+\Delta\rho^{(n)},\\
\Delta\rho^{(n)}&=&-\tfrac 1 2\nabla_{s}\,\eps^2_m\left (\rho^{(n)} \right)
,\\
\nonumber
&=&-\Ps [\bm I-\Pm] \rho^{(n)},
\end{eqnarray}
where $\nabla_{s}=\Ps \nabla$ is the component of the
gradient in the support. This algorithm is usually written as 
a projection algorithm:
\begin{eqnarray}% \nonumber
\rho^{(n+1)}%&=&\rho^{(n)}+\Delta\rho^{(n)}\,,\\
&=&\bm{P_s P_m}\rho^{(n)}\,;
\end{eqnarray}
by projecting back and forth between two sets, it converges to the
local minimum. Such algorithm is commonly referred to as Error
Reduction (ER) in the phase retrieval community \cite{Fienup:1982}.

Notice that a step of $-\half \nabla\eps^2_m(\rho)$ brings the error
$\eps^2_m(\rho)$ to 0. By projecting this step, setting to 0 some of
its components, we reduce the step length. 
Typically\cite{stark:1987,yula} the optimal step is longer
than this step, we can move along this direction and minimize further
the error metric.  The simplest acceleration strategy, the steepest
descent method, performs a line search of the local minimum in the
steepest descent direction:
\begin{eqnarray}
 \min_{\delta} \eps_m^2 \left (\rho+\delta \Delta \rho \right).
\end{eqnarray}
At a minimum any further movement in the direction of the current step
increases the error metric; the gradient direction must be
perpendicular to the current step. In the steepest descent case, where
the step is proportional to the gradient, the current step and the
next become orthogonal:
\begin{eqnarray}
\nonumber
\tfrac{\partial}{\partial \delta}\eps^2_m(\rho+\delta \Delta \rho_s)&=&
\left \langle
\Delta \rho_s |2 \Ps [\bm I-\Pm] \left (\rho+\delta \Delta \rho_s \right )
\right \rangle_r\,,\\
0&=&\left \langle
\Delta \rho_s | [\bm I-\Pm] \left (\rho+\delta \Delta \rho_s \right )
\right \rangle_r\,,
\label{eq:crossER_real}
\end{eqnarray}
where $\langle \bm x|\bm y \rangle_r=\Re \left (
\bm x^{\dagger}\cdot \bm y \right )$ and the projector $\Ps$ is
redundant and is removed.  The line search
algorithm can use $\epsilon^2_m$, and/or its derivative in
(Eq. (\ref{eq:crossER_real})). This optimization should be performed in
reciprocal space, where the modulus projector is fast to compute
(Eq. (\ref{eq:modulus})), while the support projection requires two
Fourier transforms:
\begin{eqnarray} %\tPm \trho( k)&=&\frac{\trho(k)}{|\trho( k)|}\sqrt{I( k)} \,,\\
\tPs &=&{\cal F} \bm P_s {\cal{F}}^{-1}\,,
\end{eqnarray}
 but it needs to be computed only once to calculate $\Delta \rho_s$.

The steepest descent method is known to be inefficient in the presence
of long narrow valleys, where imposing that successive steps be
perpendicular causes the algorithm to zig-zag down the valley.  This
problem is solved by the non-linear conjugate gradient method
\cite{hestenes, numrec, powell,  polak2, fletcher}. 
While error minimization  methods converge to stable solutions quite
rapidly however, the number of local minima is ofter too large for practical
ab-initio phase retrieval applications and minimization methods are
used only for final refinement.

\section{Saddle-point optimization}
The following algorithm is a reformulation of the HIO algorithm from a
gradient/constraint perspective. We seek the saddle point of the
error-metric difference ${\cal L}(\rho)=\eps^2_m(\rho)-\eps^2_s(\rho)$
\cite{unified}:
\begin{equation}
\min_{\rho_s} \max_{\rho_\us} {\cal L}(\rho_s+\rho_\us),
\label{eq:minmax}
\end{equation}
 by moving in the steepest
descent direction for $\rho_s$ ($-\Ps \nabla$) and ascent direction 
($+\Pns \nabla$)  for $\rho_\us$.  For reasons discussed in appendix (\ref{appendix:relaxation}), 
we reduce the  $\Pns$ component by a relaxation parameter
 $ \bar \beta\in[0.5,1]$:
\begin{eqnarray}
\Delta \rho^{(n)}
	&=& \{ -\Ps + \bar \beta \Pns \} \half \nabla {\cal L}\left
(\rho^{(n)} \right ).
\label{eq:hio_grad_step}
\end{eqnarray} 
The gradient of ${\cal L}$ (from Eqs. (\ref{eq:gradient_m},\ref{eq:gradient_s})):
\begin{eqnarray}
\nabla {\cal L}(\rho)&=&2[\Ps-\Pm] \rho\,,
\label{eq:minmaxgrad}
\end{eqnarray}
is used in Eq. (\ref{eq:hio_grad_step})
to express the step and the new iteration point $\rho^{(n+1)}$ as: 
\begin{eqnarray}
\nonumber
\Delta \rho^{(n)} &=&\{\Ps [\Pm -\bm I]- \bar \beta \Pns \Pm\}\rho^{(n)}\,,\\
\rho^{(n+1)}&=&
[\bm{P_s P_m+P_{\us}} [\bm{I}-\bar \beta \bm{P_m}]]\rho^{(n)}.
\label{eq:HIOrec}
\end{eqnarray}
This iteration  can be expressed in a more familiar form of the HIO algorithm
 \cite{Fienup:1978,Fienup:1982}:
\begin{eqnarray}
\label{eq:HIO}
 \rho^{(n+1)}(\bm r)&=&
\begin{cases}
\Pm \rho^{(n)}(\bm r)  & 
	\text {if  $\bm r \in S$,} \\
(\bm{I}-\bar \beta \Pm)\rho^{(n)}(\bm r)  & \text{otherwise.}
\end{cases}
\end{eqnarray}
Rather than setting to 0 the object $\rho(\bm r)$ where it is known to
be 0 ($\bm r \notin S$), this algorithm seeks a stable
condition of a feedback system in which the nonlinear operator $\Pm$
provides the  feedback term $\Pns \Pm\rho$. 
From a fixed point  whereby the feedback is 0 but 
the constraint is violated
($\Pns \rho\neq 0$) it is often possible to obtain a solution 
by a simple projection $\Pm \rho$ \cite{elser:2003}. In fact $\Pm \rho$ often satisfies
the constraints better than the current iteration $\rho$, as the
algorithm tries to escape a local minimum.

\begin{figure}
	\includegraphics[width=0.48\textwidth]{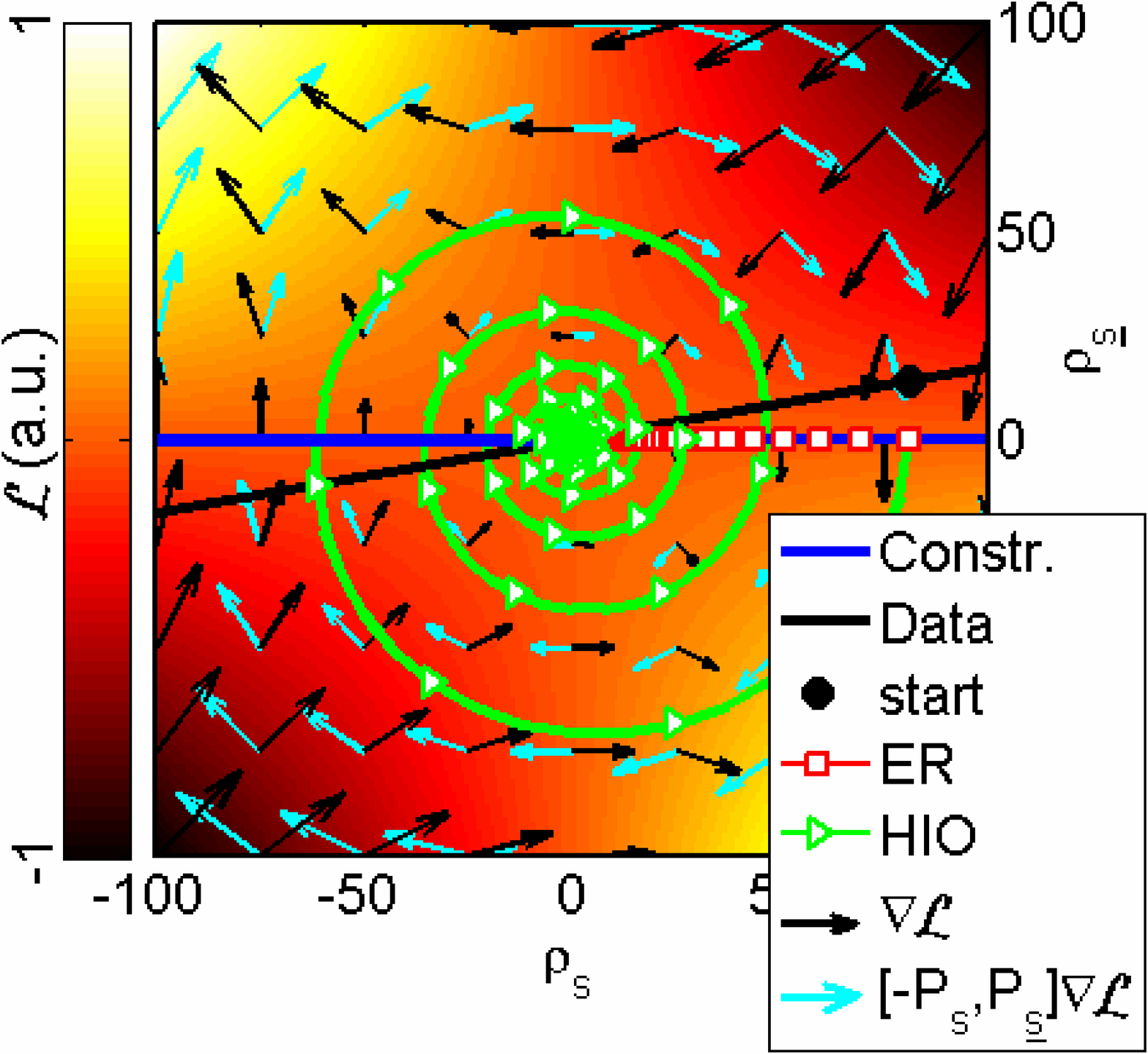}
\caption{
Algorithms seek the intersection between two sets (data and
constraints represented by two lines) using different types of fixed points: ER projects
back and forth between the two sets, moving toward the minimum of $\eps_m$ within the constraint
(horizontal line).  HIO seeks the saddle point ($\min_s \max_\us$) of
(${\cal L}$), represented in background in pseudo colormap. ${\cal L}$
is the difference of the square distances ($\eps^2$) between the
current point and the two sets.  The gradient ($\nabla {\cal L}$) is
indicated by black arrows.  In order to reach the saddle point, HIO
spirals toward the solution by inverting the gradient component
parallel to the constraint (horizontal direction) following a a
descent-ascent direction ($[-\Ps+\Pns]\nabla {\cal L}$ in light blue)
toward the solution.  Marker symbols are plotted every 5 iterations.
See also \cite{unified} for a comparison with other algorithms.
\label{2lines}}
\end{figure}

In the steepest descent method, optimization of the step length is
obtained by increasing a multiplication factor $\delta$ until the
current and next search directions become perpendicular to one
another:
\begin{eqnarray}
\left \langle
\Delta \rho|
[-\Ps +\bar \beta \Pns]\nabla {\cal L}
(\rho+\delta \Delta \rho)
\right \rangle_r&=&0\,.
\label{eq:saddle1d}
\end{eqnarray}

 A more robust strategy involves replacing the one dimensional search
 with a two dimensional optimization of the saddle point:
\begin{eqnarray}
\nonumber
\min_{\alpha}\max_{\beta} \psi(\alpha,\beta)\,,\\
\label{eq:minmaxab}
\psi(\alpha,\beta)=
{\cal L}
(\rho+\alpha \Delta\rho_s+\beta \Delta\rho_\us)\,,\\
\Delta \rho_s  =-\half \nabla_s {\cal L} (\rho);
\Delta \rho_\us=\half \nabla_\us {\cal L} (\rho);
\label{eq:steps}
\end{eqnarray}
where both components ($\Ps,\,\Pns$) of successive steps are
perpendicular to one another:
\begin{eqnarray}
\nonumber
\tfrac{\partial\psi}{\partial\alpha }&=&
 \left \langle 
\mbox{\small {$
	\Delta \rho_s|\nabla {\cal L}
	\left (
		\rho+\alpha \Delta \rho_s
		+\beta\Delta \rho_\us
	\right)
$}}
\right \rangle_r=0,\\
\tfrac{\partial\psi}{\partial\beta }&=&
 \left \langle 
\mbox{\small {$
	\Delta \rho_\us|\nabla {\cal L}
	\left (
		\rho+\alpha \Delta \rho_s+
		\beta\Delta\rho_\us
	\right)
$}}
\right \rangle_r=0.
\label{eq:perp0}
\end{eqnarray}
This two dimensional minmax problem needs to be fast to provide real
acceleration,  and will be discussed in the following section.

\section{Two dimensional subproblem \label{sec:2D}}
The local saddle point (Eq. (\ref{eq:minmaxab})) requires two conditions
to be met.  The first order condition is that the solution is a
stationary (or fixed) point, where the gradient of $\psi$ is 0 
(Eq. (\ref{eq:perp0})).  We rewrite the condition in a compact form:
\begin{eqnarray}
\label{eq:optimhio1}
\nabla_{\bm \tau} \psi(\bm \tau) =
\left \langle
\bm \Delta \rho|\bm \nabla_{\rho} {\cal L}(\rho+\bm \tau^{\mathrm{T}} \bm \Delta \rho
\right \rangle_r =0,\\
\bm  \tau=
\left (
\begin{smallmatrix}
\alpha \\ \beta
\end{smallmatrix}\right ),
\bm \Delta \rho=
\left (
\begin{smallmatrix}
\Delta \rho_s\\\Delta\rho_\us
\end{smallmatrix}\right ),
\bm \nabla_{\rho}=
\left (
\begin{smallmatrix}
\nabla_s\\ \nabla_\us
\end{smallmatrix}\right ).
\end{eqnarray}
At the origin ($\bm \tau=0$), the gradient $\nabla_{\bm \tau} \psi$  
is negative in the  $\Ps$ subspace and positive in the $\Pns$ subspace,
decreasing (increasing) ${\cal L}$ in the steepest descent (ascent)
directions in the two orthogonal spaces:
\begin{eqnarray}
\nabla_{\bm \tau} \psi (0)
&=&
\left <
\begin{smallmatrix}
 \Delta \rho_s   | \nabla_s   {\cal L}(\rho)\\
 \Delta \rho_\us | \nabla_\us {\cal L}(\rho)
\end{smallmatrix}\right >\,,\\
\nonumber
&=&
2 \left (
\begin{smallmatrix}
-||{\Delta \rho_s}||^2 \\ +||{\Delta \rho_\us}||^2
\end{smallmatrix}\right )=
\half \left (
\begin{smallmatrix}
-||{\nabla_s {\cal L}(\rho)}||^2 \\ +||{\nabla_\us {\cal L}(\rho)}||^2
\end{smallmatrix}\right ).
\end{eqnarray}
The minimal residual method finds a stationary point by minimizing the norm of the gradient:
\begin{equation}
\min_{\bm \tau} \Phi(\bm \tau),\,\, 
\Phi=\half \vectornorm{\nabla_{\bm \tau}\psi(\bm \tau)}^2\,,
\end{equation}
transforming the saddle point problem in a minimization problem, and
providing the metric $\Phi$ to monitor progress. However this method
can move to other stationary points.

Second order conditions (min-max) require the Hessian 
$\cal H$ of $\psi$ (the Jacobian of (Eq. 
(\ref{eq:optimhio1})) to be symmetric and indefinite (neither positive
nor negative definite):
\begin{equation}
{\cal H_{\tau}}=
\left (
\begin{smallmatrix}
\partial_\alpha \partial_\alpha &
\partial_\alpha  \partial_\beta \\
\partial_\beta \partial_\alpha &
\partial_\beta \partial_\beta
\end{smallmatrix}
\right )
\psi,
\begin{cases}
{\cal H}_{\alpha,\alpha}\ge 0, \\
{\cal H}_{\beta,\beta}\le 0.
\end{cases}
\label{eq:IIcond}
\end{equation}
This Hessian is computed analytically (see appendix (\ref{appendix:hessian}), 
it is small ($2\times 2$), and can be used to compute
the Newton step:
\begin{equation}
\Delta \tau=-{\cal H}^{-1} \nabla_{\tau}\psi\,.
\label{eq:newton}
\end{equation}
However, the Hessian precise value is not necessary and requires an
effort that could be avoided by other methods.

\begin{figure}[tbp]
\includegraphics[width=0.45\textwidth]{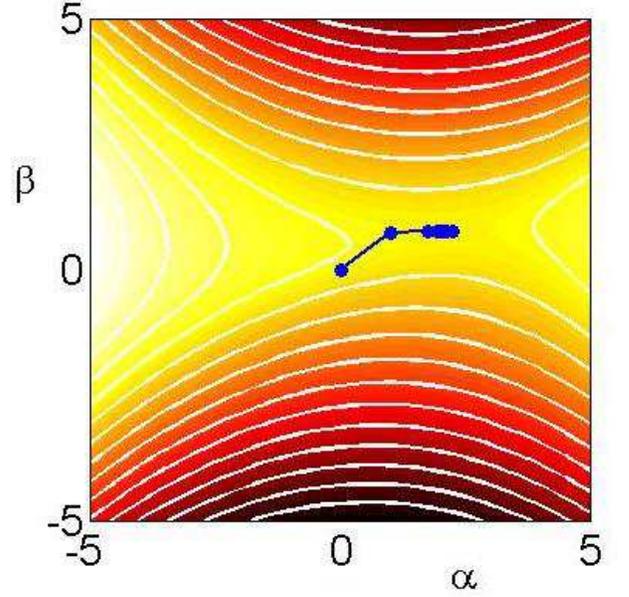}	
\caption{Pseudocolor and contour maps of the lower dimensional
function $\psi(\alpha,\beta)$ are depicted in 
the background.  This $\psi$ was computed from one iteration of the
test described in Sec. (\ref{sec:performance}).  The saddle, typical
of this test, is wide in the horizontal direction and narrow in the
vertical, showing the importance of the relaxation parameter $\bar
\beta<1$ to reduce (precondition) the vertical move. However, the
condition number increases near the saddle-point as it becomes even
narrower in the vertical direction.  Often the dependence of $\psi$
with respect to the fitting parameter $\beta$ (vertical axis)
resembled a v centered near the solution, and successive iterations
using just a preconditioner jumped up and down near the saddle-point.
Iterations using the SR1 quasi-Newton update (starting from $\bm
\tau^{(0)}=0$, ${\cal H}^{(0)}=\hat {\cal H}^{\textrm{HIO}}$) shown
here converge rapidly to the solution.}
\label{fig:saddle}
\end{figure}

The normalized steepest descent-ascent (SDA) direction can be expressed in
terms of a Newton step using an approximate diagonal Hessian whose
elements are equal to $-\nabla \psi(0)$:
\begin{eqnarray}
\hat {\cal H}^{\mathrm{SDA}} &=& 2 \left (
\begin{smallmatrix}
||\Delta \rho_s||^2  & 0\\ 0 & -||\Delta \rho_\us||^2
\end{smallmatrix}
\right ) \,.
\end{eqnarray}
The Hessians $\hat {\cal H}$ satisfies condition (Eq. (\ref{eq:IIcond})), 
ensuring that $\Delta
\tau$ is less then $90^\circ$ from the direction of the saddle.  
Starting from $\bm \tau^{(0)}=\bm 0$, the first iteration gives a unit step  
$\bm \tau^{(1)}=-\hat {\cal H}^{-1}  \nabla
\psi(\bm 0)=
\left (
\begin{smallmatrix}
1 \\ 1 
\end{smallmatrix}
\right )$. A preconditioner can be used to reduce the feedback:
\begin{eqnarray}
\hat {\cal H}^{\mathrm{HIO}} &=& \left (
\begin{smallmatrix}
1 & 0\\ 0 & 1/\bar \beta
\end{smallmatrix}
\right ) \hat {\cal H}^{\mathrm{SDA}}\,,
\label{eq:hio2d}
\end{eqnarray}
providing the HIO step at the first iteration, 
$\bm \tau^{(1)}=\Delta \bm \tau^{(0)}=
\left (
\begin{smallmatrix}
1 \\ \bar \beta 
\end{smallmatrix}
\right )$. 
This approximate Hessian can be used as a starting 
guess, which is often all it is needed to achieve fast 
convergence. 

We can perform a line search using the preconditioner $\hat {\cal H}^{-1}$:
\begin{eqnarray}
\left \langle
 \Delta \bm \tau | \hat {\cal H}^{-1} \nabla_{\bm \tau} \psi(\bm \tau+\delta \Delta \bm \tau)
\right \rangle_r=0\,.
\end{eqnarray}
However the Hessian of $\hat {\cal H}^{-1} \psi$ is antisymmetric, the
algorithm is unstable and could spiral away from the solution.  The
bi-conjugate gradient method applies to symmetric indefinite Hessians
and monitors progress of the algorithm.  Conjugate directions 
$\Lambda \bm \tau$ replace the steepest descent direction in the line
search (with $\gamma^{(0)}=0$):
\begin{eqnarray}
\Lambda
\bm \tau^{(n+1)}&=& \Delta\bm  \tau^{(n+1)}+\gamma^{(n)}\Lambda\bm \tau^{(n)}\,,\\
\nonumber
\gamma^{(n)}&=&\tfrac{
	\left \langle \Delta \tau^{(n+1)} | 
	\hat {\cal H}^{(-1)}
 \left ( \bm \nabla \psi \left(
 \tau^{(n+1)} \right ) - 
\bm \nabla \psi \left (\tau^{(n)} \right) \right ) \right \rangle
	} { \left \langle  \Delta \bm\tau^{(n)} | 
	\hat {\cal H}^{-1}
	\Delta \bm \nabla \psi \left (\tau^{(n)}\right ) \right \rangle }.
\end{eqnarray} 
A better option is to use a quasi-Newton update of the Hessian or its
inverse (a secant method in higher dimensions) based on the new
gradient values. The {\it Symmetric Rank 1} (SR1) method can be
applied to indefinite problems \cite{wright}:
\begin{eqnarray}
\nonumber
\bm y&=&
\bm \nabla_{\bm \tau} \psi(\bm \tau+\Delta \bm \tau)
-\bm \nabla_{\bm \tau} \psi(\bm \tau)\,,
\\
\nonumber
\Delta {\cal H}^{-1}&=&
\frac{
\vectornorm{\Delta \bm \tau - {\cal H}^{-1} \bm y}^2}
{(\Delta \bm \tau - {\cal H}^{-1} \bm y)^{\mathrm{T}}\cdot \bm y}\,,
\\
 {\cal H}^{-1}&\rightarrow& {\cal H}^{-1}+\Delta {\cal H}^{-1}
\,.
\label{eq:SR1}
\end{eqnarray}

  Second order conditions
(Eq. (\ref{eq:IIcond})) can be imposed to the Hessian, by flipping the
sign or setting to 0 the values that violate them. $\Phi$ can be used
to monitor progress, as long as we are in the neighborhood of the solution
and the Hessian satisfies second order conditions
(Eq. (\ref{eq:IIcond})). It was found that the Hessian and step size
 parameters where fairly constant for each 2D optimization, 
therefore the first guess for $\bm \tau$ and ${\cal H}$ was obtained
from the average of the previous 5 of such optimizations.
With such initial guess, 3 SR1 iterations  of the lower
dimensional problem where often sufficient
to reduce $\Phi$ below a threshold of 0.01 $\Phi(\bm 0)$ in the 
tests described below.

In summary, an efficient algorithm is obtained by 
a combination of HIO/quasi-Newton/Newton methods with a trust region 
$|\Delta \bm \tau|\le r$:
\begin{enumerate}
\item calculate  step 
$\bm \Delta \rho=-\half
\left (
\begin{smallmatrix}
\nabla_s {\cal L}\\-\nabla_\us {\cal L}
\end{smallmatrix}\right )$, and set trust region radius $r=r_{\max}$.
\item if the iteration number is $\le 5$, use HIO as first guess: 
${\cal H}=\hat {\cal H}^{\mathrm{HIO}}$, $\bm \tau^{(1)}=(1,\bar \beta)$.
\item otherwise average 5 previous optimized step sizes 
$\bm \tau$, and Hessians ${\cal H}$, and use the average as initial guess.
\item calculate gradient $\bm \nabla\psi(\bm \tau)$.  
If small, exit loop (go to \ref{item:updaterho}).
\item compute Newton step using approximate Hessian: 
$\Delta \bm \tau=- {\cal H}^{-1} \bm \nabla \psi$, enforce trust
region $|\Delta \bm \tau|<r$.
\item update inverse Hessian with SR1 method (Eq. \ref{eq:SR1}).
\item if the Hessian error
 $\vectornorm{\Delta \bm \tau - {\cal H}^{-1} y}^2$ is too large,
 calculate the true Hessian, perform a line search, decrease trust
 region radius $r$.
\item force Hessian to satisfy second order conditions 
(Eq. \ref{eq:IIcond}), by changing the sign of the values 
that violate conditions.
\item update $\bm \tau\rightarrow \bm \tau+ \Delta \bm \tau$ 
and go back to 4.
\item update $\rho\rightarrow\rho+\bm \tau^{\mathrm{T}} \bm \Delta \rho$. If
$\bar \eps_{m}$ is small exit, otherwise go back to 1. \label{item:updaterho}
\end{enumerate}

The trust region is used to obtain a more robust algorithm, is reset
at each inner loop, it increases if things are going well, decreases
if the iteration is having trouble, but it is kept between
($r_{\min}$, $r_{\max}$), typically $(0.5, 3)$. We can keep track of
$\bm \tau, \nabla \bm \tau$ computed, and restart the algorithm once
in a while from the root of the 2D linear fit of $\bm \nabla \psi(\bm
\tau)$.

We can easily extend this algorithm to two successive steepest 
descent-ascent directions, by performing a higher dimensional (4D) saddle-point
optimization:
\begin{equation}
\nonumber
\min_{\alpha^{(n,n+1)}} \max_{\beta^{(n,n+1)}}
{\cal L} \left ( 
\rho + (\bm \tau^{(n+1)})^{\mathrm{T}} \bm \Delta\rho^{(n+1)}
+ \mbox{\small {$(\bm \tau^{(n)})^{\mathrm{T}} \bm \Delta\rho^{(n)}$}}
\right )
\end{equation}
This 4D optimization is performed using the same Newton/quasi-Newton
trust-region optimization as in the 2D case. The first step $(\tau^{0})^{\mathrm{T}}\Delta
\rho^{0}$ is obtained solving the 2D minmax problem, and the following
ones will be perpendicular to the last 2 iterations.

\section{Performance tests \label{sec:performance}} 
\begin{figure}[tbp]
\fbox{	\includegraphics[width=0.2\textwidth]{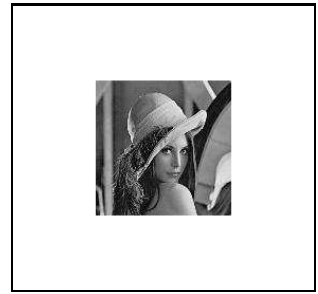}	}
\caption{Test figure used for benchmarking (total size: $256^2$,
object size: $128^2$ pixels. The support was slightly larger than
the object: $129^2$ pixels.
\label{fig:lena}
}
\end{figure}

The tests presented here were combined with other algorithms in
\cite{unified}. In such comparison between algorithms, the original
HIO algorithm performed better than other algorithms described in the
literature.  Fig. \ref{fig:lena} was used to simulate a diffraction
pattern, and several phase retrieval tests were performed using
different random starts. When applying a nonnegativity constraint HIO
always converged within a few hundred iterations. The algorithms were
therefore tested using a more difficult problem, nonnegativity and
reality constraints were removed, allowing the reconstructed image to
be complex, adding many degrees of freedom within the constraint.

 When the error metric $\eps_m$ fell below a threshold it was
considered a successful reconstruction. The threshold $10^{-4}$ chosen
was enough to obtain visually good reconstructions. 

\begin{figure}[tbp]
	\includegraphics[width=0.45\textwidth]{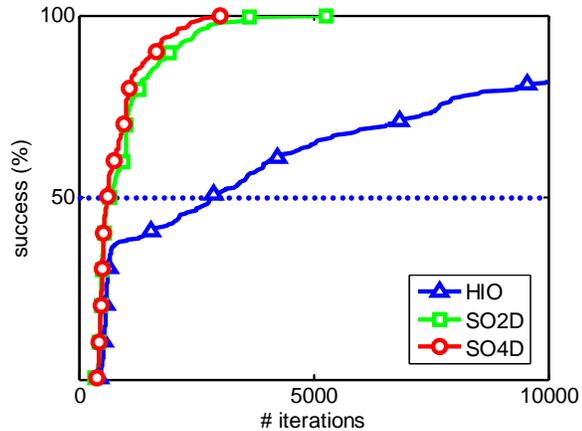}	
\caption{
Succesfull phase retrievals, starting from random phases, 
as a function of iterations. 
When the normalized error metric $\bar \eps_m$ falls below a threshold ($10^{-4}$) it is
considered a succesfull reconstruction. The plot represents the 
cumulative sum of the number of succesfull reconstructions  vs the
number of iterations required. Values for 50\% (dotted line) and 100\%
success  are listed in Tab. \ref{tab:benchmark}. Comparison with other algorithms is shown
in \cite{unified}.
\label{fig:success}
}
\end{figure}

Fig. \ref{fig:success} shows the relative performance of the various
algorithms. By adding 2D or 4D optimization, the algorithm 
converged more reliably and in less iterations to a solution 
(Table \ref{tab:benchmark}). 
The 2D (and 4D) optimization, written in an upper-level language
(matlab) increased the computational time of each iteration
by a factor of 3 (and 4) and required  the storage  
2 (and 4) additional matrices compared to HIO. HIO itself requires
 2 matrices in addition to the data and constraints.  A c-code version
of the algorithm developed by F. Maia from U. Uppsala
performed slightly faster, and could be further optimized by reducing
some of the redundant computations involved.

This lower dimensional optimization employs matrix cross products,
computing a number or floating point operations proportional to the
number of elements of the images (and can be implemented on parallel
systems). The Fourier transforms employed to calculate the steps in the
higher dimensional problem will dominate the computational burden for
larger matrices.

\begin{table}[tbp]
\caption{
Benchmark of various algorithms (250 trials) summerizing the results
in Fig. \ref{fig:success}.
 \label{tab:benchmark}
}
\begin{tabular}{|l|l|l|l|}
\hline
Algorithm &
\multicolumn{2}{|c|}{ No. of iterations  for}   & success rate  \\
          &  50\% success &100\% success &  (total)\\
\hline
HIO& 2790 &	$>10000$& 82\% 	\\
SO2D& 656 &	$5259$& 100\%	\\
SO4D& 605 &	$2999$& 100\%	\\
\hline
\end{tabular}
\end{table}

\section{Conclusions \label{sec:conclusions}}
The hybrid input-output method, usually described as a projection
algorithm, is a remarkable method to solve nonconvex phase
problems. When written as a saddle point problem, HIO speed and 
reliability can be improved by applying Newton methods to 
explore  lower dimensional search directions. This approach differs 
from other nonlinear optimization algorithms
that try to satisfy constraints using various forms of barriers or
trust regions, often requiring stochastic methods to climb out of 
local minima present in nonconvex problems. 

The saddle-point algorithm, while following a path indicated by a
gradient, does not seek a minimum and does not stop at local
minima. Although stagnation occurs, it appears that the area of
convergence to the global solution is much larger compared to simple
minimization methods.  Problems arise at the locations of 0 intensity
values where phase singularities occur, causing optimization problems
to be nonsmooth. Various forms of preconditioning low intensity values
could be applied to this problem \cite{isernia,Oszlanyi}, here the
nonsmooth behavior of the problem is addressed using a lower
dimensional optimization to stabilize the iterations, providing a more
reliable algorithm.

In this paper, the saddle-point optimization was performed for a
simple equality constraint $\Pns\rho=0$.  Such linear constraint
allows rapid calculation of the gradients involved in the lower
dimensional optimization.  Linear approximation or more advanced methods
could be applied to other nonlinear constraints such as thresholds,
histograms, atomicity and object connectivity, extending this approach
to a larger class of problems.  This saddle-point optimization
formalism can easily be generalized to other problems of conflicting
requirements where gradients or projections can be computed.

\appendix

\section{Relaxation parameter and phase singularities \label{appendix:relaxation}}
The large dimensional minmax problem (Eq. \ref{eq:minmax}) can be
expressed in a system of two parts:
\begin{equation}
\begin{cases}
\min_{\rho_s} \eps^2_m(\rho)=\min_{\rho_s} \vectornorm{[\bm I - \Pm]\rho}^2\\
\\
\min_{\rho_\us} \eps^2_s(\rho)-\eps^2_m(\rho)
=\min_{\rho_\us} 2\langle \Pm \rho|\rho\rangle_r+c
\end{cases}
\end{equation}
The upper optimization is similar to the problem treated in Section
\ref{sec:ER}, converging to a local minimum with a simple projected
gradient method. The lower function, however, can be discontinuous in
the presence of zeros ($\trho_s=0$) in Fourier space:
\begin{equation}
\langle \tPm \trho|\trho\rangle=
 \sum \sqrt{I} |\trho_s+\trho_\us|
\end{equation}
which is a non-smooth v-shaped function of $\trho_\us$ for $\trho_s=0,
\sqrt{I}>0$, and simple gradient methods oscillate around 
minima.  The projected gradient step can be overestimated and
requires the relaxation parameter $\bar \beta$.  Zeros in Fourier
space are candidates (necessary but not sufficient condition) for the
location of phase vortices, phase discontinuities, which are known to
cause stagnation \cite{fienup:stagnation}.  Analytical \cite{fiddy},
statistical \cite{fienup:stagnation, Marchesini:XRM,Miao:2006}, and
deterministic \cite{isernia,Oszlanyi} methods have been proposed to
overcome such singularities.

\section{Two dimensional gradient and Hessian \label{appendix:hessian}}
The function ${\cal L}$ in reciprocal space can be expressed as:
\begin{eqnarray}
{\cal L}(\trho_s+\trho_\us)&=&
\vectornorm{[\bm I-\tPm] (\trho_s+\trho_\us)}^2-
\vectornorm{\trho_\us}^2\\
\nonumber
&=&\sum |\trho_s|^2-2\sqrt{I}|\trho_s+\trho_\us|+\sqrt{I}\,.
\end{eqnarray}
and the two components of the gradients:
\begin{eqnarray}
\bm \nabla {\cal L}&=&
 \left (
\begin{smallmatrix}
\Ps \nabla {\cal L}\\
\Pns\nabla {\cal L}
\end{smallmatrix}
\right )
=
2\left (
\begin{smallmatrix}
\Ps [\bm I-\Pm](\rho_s+\rho_\us)\\
-\Pns \Pm(\rho_s+\rho_\us)
\end{smallmatrix}
\right )
\end{eqnarray}
The corresponding steps 
$\Delta \rho_s=-\half \nabla_s {\cal L}$,
$\Delta \rho_\us=+\half \nabla_\us {\cal L}$.

The function $\psi(\alpha,\beta)$ can be calculated in reciprocal
space, provided that the components $\trho_{s,\us},\Delta\trho_{s,\us}$
are known:
\begin{eqnarray}
\nonumber
\psi(\alpha,\beta)&=&
\vectornorm{[\bm I -\tPm]
\left (
	\trho+\alpha \Delta \trho_s
	+\beta \Delta \trho_\us
\right)
}^2\\
\nonumber
&-&\vectornorm{\trho_\us+\beta \Delta\trho_\us }^2
\\
\nonumber
&=&\sum_{\bm k}
\left |
	\left |
	\trho+\alpha \Delta \trho_s
	+\beta \Delta \trho_\us
	\right |-\sqrt{I}
\right |^2\\
\nonumber
 &&-
\left |
\trho_\us+\beta \Delta \trho_\us
\right |^2\\
&=& 
\sum_{\bm k} I+
\left |
\trho_s+\alpha \Delta \trho_s
\right |^2\\
\nonumber
&&-2 \sqrt{I}
\left |
\trho+\alpha \Delta \trho_s
+\beta \Delta \trho_\us
\right |
\end{eqnarray}
The gradient components 
(writing $\trho_\tau=\trho+\bm \tau^{\mathrm{T}}\bm \Delta \trho=
\trho+\alpha\Delta\trho_s+\beta\Delta\trho_\us$)
are:
\begin{eqnarray}
\mbox{\small {$
\bm \nabla_{t} \psi
$}}
&=&
 \left \langle
\mbox{\small {$
	\bm \Delta \trho |   \nabla {\cal L}( \trho_\tau)
	$}}
\right \rangle_r\,,
\\
\nonumber
&=&2 \left \langle
\mbox{\small {$
	\bm \Delta \trho | [\tPs-\tPm]
		\trho_\tau
	$}}
\right \rangle_r\,,
\\
\nonumber
\tfrac{\partial\psi}{\partial\alpha }&=&
2 \left \langle 
\mbox{\small {$
	\Delta \trho_s|[\bm{I}-\tPm]
		\trho_\tau
$}}
\right \rangle_r\,,\\
\nonumber
\tfrac{\partial\psi}{\partial\beta }&=&
2 \left \langle 
\mbox{\small {$
	\Delta \trho_\us|[-\tPm]
		\trho_\tau
$}}
\right \rangle_r\,,
\end{eqnarray}

Using common derivative rules:
\begin{eqnarray}
\tfrac{\partial} {\partial x} |x|&=&\tfrac {x}{|x|},\\
\tfrac{\partial}{\partial\alpha} |x+\alpha \Delta x|&=&\tfrac
{\Re\left (\Delta x^{\dagger}(x+\alpha \Delta x)\right )}
{|x+\alpha \Delta x|},\\
\tfrac{\partial}{\partial\alpha} 
\tfrac
{x+\alpha \Delta x}{|x+\alpha \Delta x|}
&=&\tfrac
{\Delta x}
{|x+\alpha \Delta x|}
-
\tfrac
{x+\alpha \Delta x }
{|x+\alpha \Delta x|}\\
\nonumber
&\cdot& \tfrac
{\Re\left (\Delta x^{\dagger}(x+\alpha \Delta x)\right )}
{|x+\alpha \Delta x|^2},\\
\tfrac{\partial}{\partial\alpha} |x+\alpha \Delta x|^2&=&
2\Re\left (\Delta x^{\dagger}(x+\alpha \Delta x)\right ),\\
\tfrac{\partial^2}{\partial\alpha^2} |x+\alpha \Delta x|&=&
\tfrac
{|\Delta x|^2}{|x+\alpha \Delta x|}
-\tfrac
{\Re \left (\Delta x^{\dagger}(x+\alpha \Delta x) \right )^2}{|x+\alpha \Delta x|^3},\\
\tfrac{\partial^2}{\partial\alpha^2} |x+\alpha \Delta x|^2&=&
\left |\Delta x \right |^2,
\end{eqnarray}
and
\begin{eqnarray}
\tfrac{\partial^2|x+\alpha \Delta x+\beta \Delta y|}{\partial \alpha\partial \beta} 
&=&
\tfrac
{\Re(\Delta x^{\dagger}\Delta y)}
{|x+\alpha \Delta x+\beta \Delta y|}\\
\nonumber
&-&\tfrac
{
\Re \left (\Delta x^{\dagger}(x+\alpha \Delta x+\beta \Delta y) \right )
}
{|x+\alpha \Delta x+\beta \Delta y|}\\
\nonumber
&\cdot& 
\tfrac
{
\Re \left (\Delta y^{\dagger}(x+\alpha \Delta x+\beta \Delta y) \right )
}
{|x+\alpha \Delta x+\beta \Delta y|^2},
\end{eqnarray}
we can calculate the analytic expression for the Hessian using $\tPm \trho= \tfrac {\trho}{|\trho|}\sqrt{I}$.
Starting from the simplest component:
\begin{eqnarray}
\tfrac{\partial^2\psi}{\partial\beta^2 }&=&
-2\left \langle
\Delta \trho_\us|
\tfrac{\partial \tPm \trho_\tau}{\partial\beta }
\right \rangle\\
\nonumber
&=&2\sum -
\tfrac{
|\Delta \trho_\us|^2
\sqrt{I}
}
{|\trho_\tau|}
+\tfrac{
\Re 
\left (
\Delta \trho_\us^{\dagger}
\trho_\tau
\right)^2 \sqrt{I}
}
{|\trho_\tau|^3}\\
\nonumber
&=&2\sum -
\tfrac{|\Delta \trho_\us|^2 \sqrt{I}}
{|\trho_\tau|}
+\tfrac{\sqrt{I}}
	{2\left |\trho_\tau \right |}
\tfrac{
\Re \left (
\Delta \trho_\us^{\dagger} 
\trho_{\tau} \right ) 
\left (
\Delta \trho_\us^{\dagger} \trho_\tau+
\Delta \trho_\us \trho_\tau^{\dagger}
\right )
}{\left |\trho_\tau \right |^2}
\\ \nonumber
&=&2
\left \langle 
\Delta \trho_\us|
-\tfrac{\sqrt{I}}{2| \trho_\tau|}
\left (1-
\tfrac{ \trho_\tau^2}{\Delta\trho_\us^2}
\tfrac{ |\Delta\trho_\us|^2}
{|\trho_\tau|^2}
\right )|
\Delta \trho_\us\right \rangle_r
\end{eqnarray}

\begin{eqnarray}
\tfrac{\partial^2\psi}{\partial\alpha^2 }&=&
2 \left \langle 
\Delta \trho_s \big \lvert
\tfrac{\partial [\bm I-\tPm]\trho_\tau}
{\partial\alpha }
\right \rangle_r \,,\\
\nonumber
&=&
2\sum
|\Delta \trho_s|^2
-\tfrac{|\Delta \trho_s|^2\sqrt{I}}
	{|\trho_\tau|}
+\tfrac{
\Re \left (
\Delta \trho_s^{\dagger}
\trho_\tau
\right)^2\sqrt{I}
}
{|\trho_\tau|^3}
\\ \nonumber &=&
2\left \langle 
\Delta \trho_s
\left \lvert
\left [
1-\tfrac{\sqrt{I}}{2\left | \trho_\tau \right|}
\left (
1-
\tfrac{\trho_\tau^2 }{\Delta\trho_s^2}
\tfrac{|\Delta\trho_s|^2}{|\trho_\tau |^2}
\right )
\right ]
\right \rvert
\Delta \trho_s\right \rangle_r
\end{eqnarray}
The cross terms:
\begin{eqnarray}
\tfrac{\partial^2\psi}{\partial\beta\partial \alpha }
&=&2
\left \langle
\Delta \trho_s \Big \lvert
\tfrac{\partial
[\bm I-\tPm]\trho_\tau}{\partial\beta }
\right \rangle
\\ \nonumber &=&
2 \left \langle 
\Delta \trho_s 
\left \lvert
-\tfrac{\sqrt{I}}{2 | \trho_\tau |}
\left (
	1-
	\tfrac{\trho_\tau^2 }{\Delta\trho_\us^2}
	\tfrac{|\Delta\trho_\us|^2}{|\trho_\tau |^2}
\right )
\right \rvert
\Delta \trho_\us\right \rangle_r
\\ \nonumber &=&
2\left \langle 
\Delta \trho_\us
\left \lvert
-\tfrac{\sqrt{I}}{2 | \trho_\tau |}
\left (
	1-
	\tfrac{\trho_\tau^2 }{\Delta\trho_s^2}
	\tfrac{|\Delta\trho_s|^2}{|\trho_\tau |^2}
\right )
\right \rvert
\Delta \trho_s\right \rangle_r \,.
\end{eqnarray}

\begin{acknowledgments}
This work was performed under the auspices of the U.S. Department of
Energy by the Lawrence Livermore National Laboratory under Contract
No. W-7405-ENG-48 and the Director, Office of Energy Research.  This
work was partially funded by the National Science Foundation through
the Center for Biophotonics, University of California, Davis, under 
Cooperative Agreement No. PHY0120999. F. Maia from Univ. of Uppsala
for porting the code in c.
\end{acknowledgments}

%\listoffigures
\end{document}